\documentclass{article}
\usepackage[T1]{fontenc}
\usepackage[latin1]{inputenc}
\usepackage{a4wide}
\makeatletter

\providecommand{\LyX}{L\kern-.1667em\lower.25em\hbox{Y}\kern-.125emX\@}

 \newcommand{\lyxaddress}[1]{
   \par {\raggedright #1 
   \vspace{1.4em}
   \noindent\par}
 }

\makeatother

\begin{document}

\title{Distinguishability of the Bell states}

\author{Sibasish Ghosh\( ^{\star } \)\thanks{
res9603@isical.ac.in
}, Guruprasad Kar\( ^{\star } \)\thanks{
gkar@isical.ac.in
}, Anirban Roy\( ^{\star } \)\thanks{
res9708@isical.ac.in
}, Aditi Sen(De)\( ^{\natural } \)\thanks{
aditisendein@yahoo.co.in
} \\
and Ujjwal Sen\( ^{\natural } \)\thanks{
ujjwalsen@yahoo.co.in
} }

\maketitle

\lyxaddress{\( ^{\star } \)Physics and Applied Mathematics Unit, Indian Statistical Institute,
203 B.T. Road, Kolkata 700 035, India }

\lyxaddress{\( ^{\natural } \)Department of Physics, Bose Institute, 93/1 A.P.C. Road,
Kolkata 700 009, India}

\begin{abstract}
More than two multipartite orthogonal states cannot always be discriminated
(with certainty) if only local operations and classical communication (LOCC)
are allowed. Using an existing inequality among the measures of entanglement,
we show that any three Bell states cannot be discriminated by LOCC. Exploiting
the inequality, we calculate the distillable entanglement of a certain class
of \( 4\otimes 4 \) mixed states. 
\end{abstract}
Nonorthogonal states cannot be discriminated with certainty. This is essentially
the No-cloning theorem \cite{1}. However discrimination with certainty is not
guaranteed even for multipartite orthogonal states, if only local operations
and classical communication (LOCC) are allowed. Given a set of multipartite
orthogonal states, at present it is not always possible (without further study)
to say whether they can be discriminated (with certainty) or not, where only
LOCC are allowed. 

A landmark result was obtained in \cite{2}, where it was shown that any two
multipartite orthogonal states can be distinguished with certainty by LOCC.
Recently Virmani et al. \cite{3} have shown that for inconclusive discrimination
of any two multipartite pure nonorthogonal states, the optimal probability of error can be attained
by using LOCC only. Chen and Yang \cite{4} have shown that the same is true  even in the case of conclusive discrimination.
In this paper we solve the question of distinguishability (with certainty) of
any set consisting of the Bell states. It was noted in \cite{2} that if \emph{two
copies} of a state are provided which is known to be one of the four orthogonal Bell states
\[
\left| B_{1}\right\rangle =\frac{1}{\sqrt{2}}\left( \left| 00\right\rangle + \left| 11\right\rangle \right), \]
\[ \left| B_{2}\right\rangle =\frac{1}{\sqrt{2}}\left( \left| 00\right\rangle - \left| 11\right\rangle \right), \]
\[ \left| B_{3}\right\rangle =\frac{1}{\sqrt{2}}\left( \left| 01\right\rangle + \left| 10\right\rangle \right), \]
\[ \left| B_{4}\right\rangle =\frac{1}{\sqrt{2}}\left( \left| 01\right\rangle - \left| 10\right\rangle \right), \]
one can discriminate between them using LOCC only. Analysing the properties
of the unlockable bound entangled state \( \frac{1}{4}\sum ^{4}_{i=1}P\left[ \left| B_{i}\right\rangle \left| B_{i}\right\rangle \right]  \)
\cite{5}, it follows that this is not possible if only a \emph{single copy}
is provided. We show that it is not possible to discriminate between any three
Bell states if only a single copy is provided and if only LOCC are allowed.
That two Bell states can be distinguished follows from the result in \cite{2}.
Throughout the paper we explicitly consider discrimination with certainty. 

Suppose now that it is possible to discriminate (with certainty) between the
four Bell states using only LOCC. Assume that there is a four-party state 

\[
\rho ^{S}=\frac{1}{4}\sum ^{4}_{i=1}P\left[ \left| B_{i}\right\rangle _{AB}\left| B_{i}\right\rangle _{CD}\right] \]
shared between Alice, Bob, Claire and Danny \cite{5}. If the four Bell states
are locally distinguishable, Alice and Bob would be able to discriminate between
them without meeting. A classical communication would then result in Claire
and Danny sharing \( 1 \) ebit. As Alice and Bob did not meet, this would contradict
the fact that \( \rho ^{S} \) is separable in the AC : BD cut as it can be
written as \( \frac{1}{4}\sum ^{4}_{i=1}P\left[ \left| B_{i}\right\rangle _{AC}\left| B_{i}\right\rangle _{BD}\right]  \).
This proves that it is not possible to discriminate (with certainty) between
the four Bell states, when only a single copy is provided. 

We now proceed to prove this result for three Bell states.

The relative entropy of entanglement for a bipartite quantum state \( \sigma  \)
is defined by \cite{6}

\[
E_{r}(\sigma )=\min _{\rho \in D}S(\sigma \parallel \rho )\]

where \( D \) is the set of all separable states on the Hilbert space on which
\( \sigma  \) is defined, and \( S(\sigma \parallel \rho )\equiv tr\{\sigma (\log _{2}\sigma -\log _{2}\rho )\} \)
is the relative entropy of \( \sigma  \) with respect to \( \rho  \). 

Consider now the state \[
\rho ^{(3)}=\frac{1}{3}\sum ^{3}_{i=1}P\left[ \left| B_{i}\right\rangle _{AB}\left| B_{i}\right\rangle _{CD}\right] \]
 where the Bell states involved are \emph{any} three of the four. Let \( E_{r}(\rho _{AC:BD}^{(3)}) \)
denote the relative entropy of entanglement of the state \( \rho ^{(3)} \)
in the AC : BD cut. Then\[
E_{r}(\rho _{AC:BD}^{(3)})\leq S\left( \rho _{AC:BD}^{(3)}\parallel \frac{1}{4}\sum ^{4}_{i=1}P\left[ \left| B_{i}\right\rangle _{AC}\left| B_{i}\right\rangle _{BD}\right] \right) =2-\log _{2}3<1\]
But distillable entanglement is bounded above by \( E_{r} \) \cite{7,8}. Consequently
the distillable entanglement of \( \rho ^{(3)} \), in the AC : BD cut, is strictly
less than unity \cite{9}. 

Suppose now that it is possible to discriminate (with certainty) between any
three Bell states when only LOCC are allowed. So, if Alice, Bob, Claire and
Danny share the state \( \rho ^{(3)}, \) then Alice and Bob, without meeting,
would again be able to make Claire and Danny share \( 1 \) ebit of distillable
entanglement. And again we reach at a contradiction. Therefore even three Bell
states are not locally distinguishable with certainty. Here we remark that the
state \( \rho ^{(3)} \) can also be called `unlockable' \cite{5} \emph{in
the sense} that it would not be possible to produce \( 1 \) ebit between C
and D when A and B are separated although \( 1 \) ebit can be generated between
C and D when A and B comes together. 

We now show that the above method can be employed to calculate the distillable
entanglement, in the AC : BD (or AD : BC) cut, of any state of the form \[
\rho ^{\prime (2)}=\frac{1}{2}P\left[ (a\left| 00\right\rangle +b\left| 11\right\rangle )_{AB}\left| B_{1}\right\rangle _{CD}\right] +\frac{1}{2}P\left[ (\overline{b}\left| 00\right\rangle -\overline{a}\left| 11\right\rangle )_{AB}\left| B_{2}\right\rangle _{CD}\right] \]
where \( \left| a\right| ^{2}+\left| b\right| ^{2}=1 \). As two orthogonal
states can always be locally discriminated \cite{2}, one can produce \( 1 \)
ebit between C and D by operating locally on A and B. Thus the distillable entanglement
of the above state would at least be \( 1 \) in the AC : BD as well as the
AD : BC cut. But the relative entropy of entanglement (which is an upper bound
of the distillable entanglement \cite{7,8}) of \( \rho ^{\prime (2)} \) is
\( 1 \) in these cuts \cite{11}. Thus the distillable entanglement of the above
state in the AC : BD (and AD : BC) cut is \( 1 \). In particular, the relative
entropy of entanglement as well as the distillable entanglement of the state
\[
\rho ^{(2)}=\frac{1}{2}\sum ^{2}_{i=1}P\left[ \left| B_{i}\right\rangle _{AB}\left| B_{i}\right\rangle _{CD}\right] \]
 is \( 1 \) in the AC : BD (and AD : BC) cut. 

Let us now summarize the results. The fact that the state \( \frac{1}{4}\sum ^{4}_{i=1}P\left[ \left| B_{i}\right\rangle \left| B_{i}\right\rangle \right]  \)
is separable in all \( 4\otimes 4 \) cuts, implies that if only a single copy is provided
and only LOCC are allowed, it is not possible to discriminate with certainty
among the four Bell states. We show that even (any) \emph{three} Bell states
cannot be distinguished by LOCC, when only a single copy is provided. We prove
this by using an existing inequality of the entanglement measures. This inequality
also helped us to obtain the distillable entanglement of a certain class of
\( 4\otimes 4 \) mixed states.

We thank Debasis Sarkar for helpful discussions. U.S. acknowledges partial support
by the Council of Scientific and Industrial Research, Government of India, New
Delhi.

\end{document}